# CERTIFICATE-BASED SINGLE SIGN-ON MECHANISM FOR MULTI-PLATFORM DISTRIBUTED SYSTEMS

Magyari Attila*, Genge Bela**, Haller Piroska**

*"Petru Maior" University of Tirgu Mures*
*Nicolae Iorga Str., No. 1, Mures (540088), ROMANIA*
*\* (Phone: +40742577702; e-mail: atti86@gmail.com).*
*\*\* (Phone: +40 265 262275; e-mail: {bgenge,phaller}@engineering.upm.ro).*

**Abstract:** We propose a certificate-based single sign-on mechanism in distributed systems. The proposed security protocols and authentication mechanisms are integrated in a middleware. The novelty of our middleware lies on the use of XPCOM components, this way we provide a different services that can be used on every platform where Mozilla is available. The component based architecture of the implemented services allows using the authentication components separately.

## 1. Introduction

In this paper we propose a single sign-on mechanism based on certificates generated on request for client applications. Single sign-on mechanisms ensure the use of user credentials for accessing multiple resources where the user is requested to enter its credentials only once. This ensures a reduction of the number of passwords used which can significantly improve security of systems by minimizing the likelihood of a password being compromised [1]. Communication between client applications and servers is done using secure channels based on security protocols. In order to minimize the overhead needed for accessing multiple servers, instead of using protocols such as SSL [2] or its more recent version TLS [3], we designed a set of new protocols based on Guttman's authentication tests [4, 5]. The protocols have been implemented using the existing security library OpenSSL [6], which, together with the protocol descriptions, ensures the correct implementation of the designed protocols.

In order to provide a minimal effort for developing single sign-on mechanisms in distributed systems, we developed a middleware that implements the proposed security protocols and single sign-on mechanism. Existing single sign-on mechanisms are either implemented to function on a single platform, such as Active Directory [7] for Microsoft Windows or eDirectory [8] for Unix systems, or they rely on a centralized directory structure such as LDAP [9], to which servers must be connected in order to authenticate users. The novelty of our middleware lies on the use of XPCOM [10] components provided by the Mozilla platform to encapsulate the communication layer. This way, we do not only provide a single sign-on mechanism for a single platform, but a mechanism that can be used on every platform where Mozilla is available.

The rest of the paper is structured as follows: In the next section we describe the architecture of the middleware: the requirements, the software stack and the security protocols.

## 2. Middleware architecture

*2.1 Requirements*

Network users typically maintain a set of authentication credentials (usually a username/password pair) with every Service Provider (SP) they are registered with. In the context of this paper a service provider is any entity that provides some kind of service or content to a user. Examples of SPs include web services, messenger services, FTP/web sites, and streaming media providers. The number of such SPs with which users usually interact has grown beyond the point at which most users can memorize the required credentials. The most common solution is for users to use the same password with every SP with which they register — a tradeoff between security and usability in favor of the latter. A solution for this security issue is Single Sign-On (SSO), a technique whereby the user authenticates him/herself only once and is automatically logged into SPs as necessary, without requiring further manual interaction [11].

There are several approaches to create a SSO network: The Kerberos based [12] systems initially prompt the user for credentials, emitting a Kerberos ticket-granting ticket (TGT). Drawbacks to the Kerberos based system include the centralized architecture: when the Kerberos server is down, no one can log in. Kerberos requires the clocks of the involved hosts to be synchronized, the tickets have a time availability period, by default configuration 10 minutes, and if the host clock is not synchronized with the Kerberos server clock, the authentication will fail. Also, the secret keys for all users are stored on the central server, so a compromise of that server will compromise all users' secret keys. Another approach would be the smart card based authentication: an integrated circuit, which can process data, is embedded in a plastic card, which will be used to identify its owner. The necessity of this hardware, which can easily be physically damaged, stolen or compromised, excluded this method from our list. Some other possibilities include the use of one-time passwords (OTP) or the integrated windows authentication but for our model we have chosen a client certificate based configuration. First of all, the X.509 certificates we're using are ITU-T standardized, which widens the possibilities of the implementations or further developing. These certificates are based on the RSA encryption algorithm, providing the necessary security needed. The certificates are relatively easily generated and due to their small size, their storage and transport over the network is also easy. The X.509 certificates store several predefined information about their owner, but can also contain custom data. We use these fields to store each client's permissions in the network. An immediate disadvantage of such an approach is the support for a single encryption algorithm at a time. It was shown that the algorithm can be broken if there are enough resources used, but using larger keys (1024 or 2048 bit) makes this very hard, if not impossible, with existing technologies. Another drawback to RSA encryption is its processing power and execution time, compared to other algorithms, like: AES, 3DES, Blowfish or RC6. This is why we try to minimize its usage, and when possible, replace it with a more resource-friendly encryption algorithm.

Single sign-on mechanisms already exist, and they are widely used, like the mentioned Active Directory for Microsoft Windows or eDirectory for UNIX systems. However they are platform-specific. Our goal was to create a mechanism that runs on a wide variety of platforms, hence we have chosen XPCOM. It stands for Cross Platform Component Object Model, and it's a framework for writing multi-platform, modular software. The core of the components is written using the NSPR (Netscape Portable Runtime [13]) libraries, as shown in *Figure 1* [14]. As an application, it uses a set of core XPCOM libraries to selectively load and manipulate XPCOM components. It's open source, and it supports just about any platform that hosts a C++ compiler, including: Microsoft Windows, Linux, HP-UX, AIX, Solaris, OpenVMS, MacOS, and BSD.

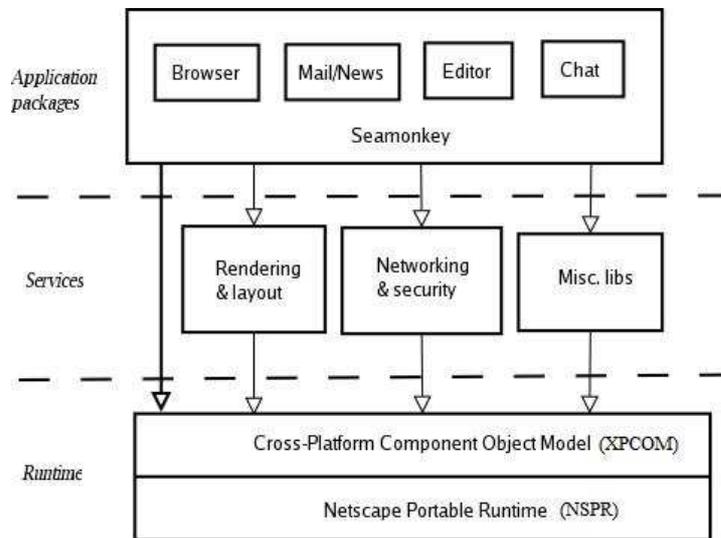

*Figure 1.* Top Level Conceptual Architecture of Mozilla Application Suite

*2.2 Software Stack*

The middleware structure has four layers, as shown in *Figure 2*.

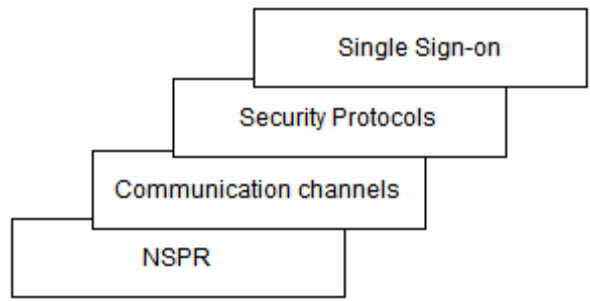

*Figure 2.* Middleware structure

*2.2.1 NSPR*

The NSPR layer of the middleware is implemented using of various classes and objects, such as threads, sockets, coders, parsers, timers, several data structures, and other implementations, which form the foundation of the whole platform. These components were written using the NSPR libraries. Netscape Portable Runtime (NSPR) provides platform independence for non-GUI operating system facilities. These facilities include threads, thread synchronization, normal file and network I/O, interval timing and calendar time, basic memory management and shared library linking. The current implementation supports Macintosh (PPC), WIN-32 (WinNT, Win9x) and 20 versions of UNIX and is still expanding.

*2.2.2 Communication Channels*

The communication channels are built on top of the NSPR layer to create more advanced data transportation mechanisms. The channels are created dynamically and managed by channel handlers. They support custom, predefined structured messages, but also raw data.

*2.2.3 Single Sign-on*

Single sign-on (SSO) is a mechanism whereby a single action of user authentication and authorization allows accessing to all computers and systems where authorization rights have been verified, without the need to enter multiple passwords. Single sign-on reduces human error, a major component of systems failure and is therefore highly desirable.

Our proposed system is composed of two types of participants: clients and servers. In *Figure 3* we illustrated a simple network with 3 servers and two clients: one already connected and another who's in the authentication process. The communication lines between the nodes may be unstable and in most of the cases unsafe, which exposes our messages to different threats like spoofing, replicating or simple message loss. We designed the system to prevent any of these attacks, and to be easy to implement and use. Each server can hosts many and different services, but for our model we only need an authentication service and a resource service. The services are of type request-response, and all the data sent is confidential. The authentication service provides two types of authentication mechanisms: the first one requires the use of a username and password, while the second one requires the use of the generated certificates. In order to gain access to a Service Provider (SP), a client first has to register at one server called the *home server*. Each server can be a home server and *resource server* at the same time; it's only relative to the client. The registration can take any form, in our model we assume that there is a secure database, where every client is already registered. The requester contacts its home server, and sends him his credentials (Step 1 in *Figure 3*); this is the only time he has to manually log in. The home server will generate a certificate, containing user data (e.g. username, location, organization name, e-mail address), expiration date, but also information about the issuer, to verify its genuineness. The certificate also contains information about the user's permissions, following a role-based access control (RBAC) model. Since users are not assigned permissions directly, but only acquire them through their role (or roles), management of individual user rights becomes a matter of simply assigning appropriate roles to the user. This simplifies common operations, such as adding a user, or changing a user's department. In Step 2 (*Figure 3*), the client receives the certificate. The next two steps, 3 and 4 in *Figure 3* are to contact the desired SP, sending the certificate, and exchanging a session key, which will be used to encrypt data from now on. RSA encryption algorithms, which we used so far, require more processing power, so we will use the triple DES algorithm, with a new key each session to maximize security and performance. If the client wants to access a different SP, it just has to send the certificate, and a new session key will be generated. As long as the certificate isn't expired, it can connect to every SP in the network, otherwise it will have to repeat the first step and obtain a new certificate.

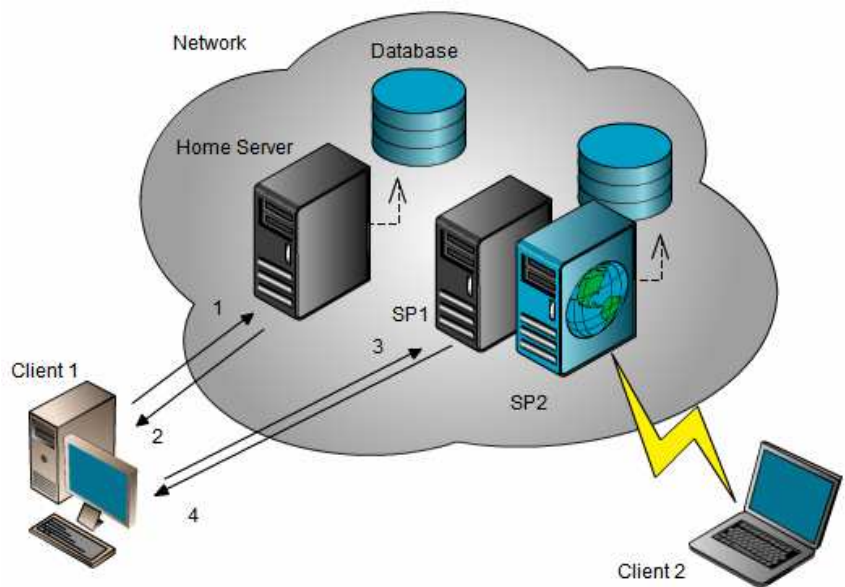

*Figure 3.* System setup

*2.3 Security Protocols*

In the proposed middleware, there was a need for authentication protocols that satisfied security requirements, such as: confidentiality in an insecure environment, supporting message loss, certificate and key generation. We developed several security protocols, based on Guttman's authentication tests. The implementation of these protocols was done using the OpenSSL security libraries. A combination of symmetric and asymmetric cryptographic algorithms was used to achieve a balance between security and performance. The authentication consists of two phases: acquiring the certificate from the home server, and the second to authenticate at the resource server with the newly generated credentials.

In order to achieve a valid certificate and key, the client (A) needs to contact its home server (B). This is where the first phase of the authentication protocol takes place (*Figure 4*), initiated by the client who sends its username, requesting a connection. If the server finds the username in its database, and the system is capable of accepting a new connection, it generates a 1024 bit length nonce (N, random). A hash function (h) is applied on this nonce, and is sent to the client, together with a message informing the other participant that it can continue with the next step. Then, the client sends its username and password, and a single secret key is generated ($K_{AB}$) used to encrypt the next message from the server. The received hash of the nonce is hashed again, and together with the username, password and the generated symmetric key, are encrypted using the server's public key ($pk_B$). Upon receiving the data from the client, the server hashes the nonce once again and compares it to the previously saved data. If they match, meaning the message is fresh, it verifies the username and password and a new certificate will be generated, along with the RSA inverse keys. The secret key ($sk_A$) will be encrypted with the key received from the client. The keys, the certificate and the nonce are digitally signed, and sent back to the client. This will verify the nonce and the signature, and if everything is valid, the certificate and the secret key are decoded and decrypted, finalizing the first phase of the authentication.

The second phase of the authentication, which you can see in *Figure 4,* starts after acquiring a certificate. The client contacts the desired resource server, communicating his intentions on getting access to the resources. If the server is willing to accept new connections, it will generate and send a 1024 bit nonce (N), informing the client about the connection being accepted. Receiving this message, the client hashes and signs the received nonce with his own private key ($sk_A$), and attaches the certificate to the message. The server can verify the signed nonce with the received certificate, but this certificate will also be verified to ensure it was emitted by a trusted authority, in this case, the client's home server. If no problems occur, the server proceeds to generate a session key ($K_{AB}$), which will be used for further data encryption. This key and the nonce will be encrypted with the client's public key ($pk_A$), and also signed by the server, to protect its contents. The whole message is encrypted again with the server's public key, to prevent any modifications on the data.

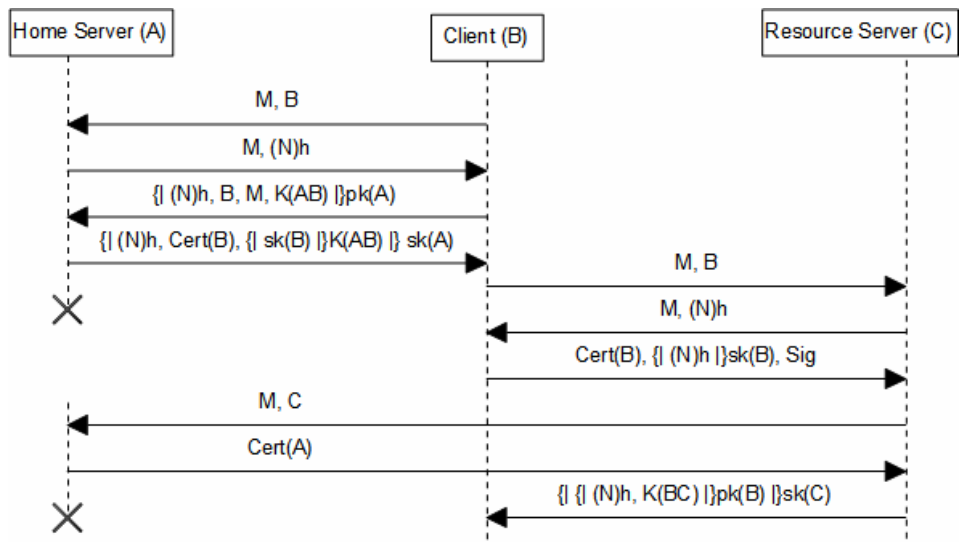

*Figure 4.* Authentication protocol

## 3. Experimental Results

The tests were performed on a Microsoft Windows machine, 2800 Mhz dual core CPU. As you can see in *Figure 6*, the RSA key generations use the most resources. When the number of clients is lower then 10, the delay could vary between 50 to 500 milliseconds, but if more than 10 clients try to request certificates simultaneously, the waiting time can go over 1-2 seconds, as you can see in *Figure 7*. This wouldn't be a problem, but in a populated network, we can't limit the number of clients to 10, there could be hundreds of even thousands of requests at the same time, and could create a bottleneck in the servers. The key generating time is directly proportional with the processing power of the CPU, so upgrading our hardware can speed up the acquiring process. There are several other ways to improve the overall performance of the system:

- Using a dedicated processor for RSA key generation, optimized only for this algorithm;
- Developing a new, or improving the current library algorithm;
- Introducing a new type of server in our system, this could analyze each server's load and balance the system by sending clients to less busy servers.

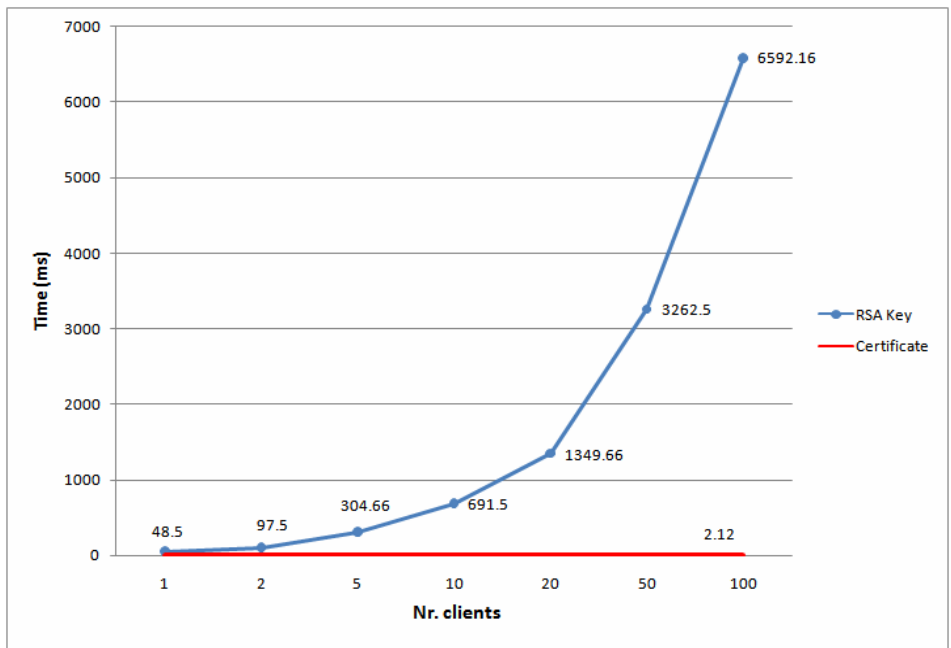

*Figure 5.* Certificate and RSA key generation in time

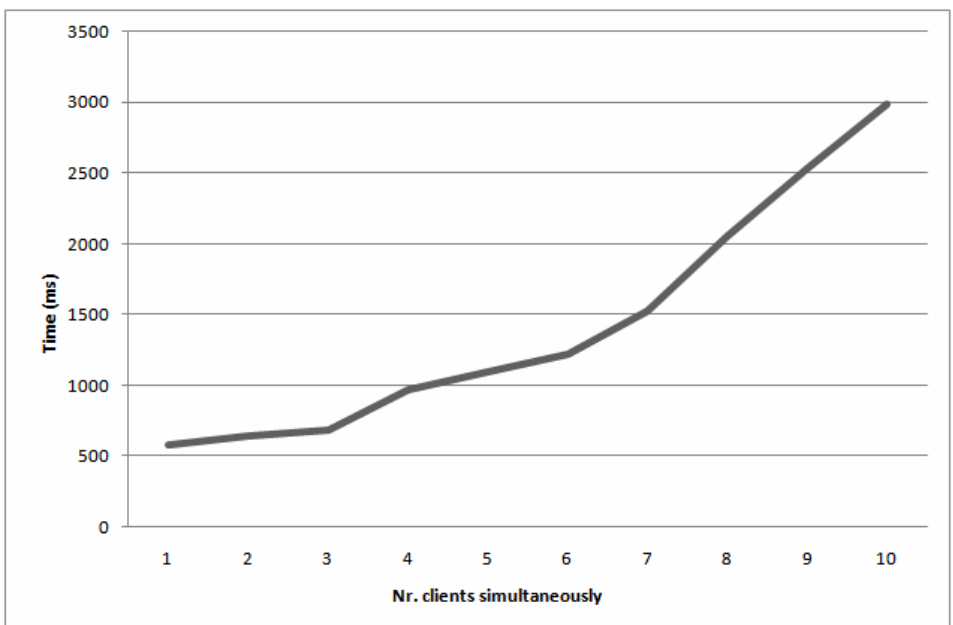

*Figure 6.* Certificate acquiring time by client

## 4. Conclusions

We implement a middleware platform based on XPCOM components to assure different services for platform independent distributed application. The proposed authentication protocol as part of the middleware was design to work in an insecure environment, supporting message loss, certificate and key generation. The implemented protocols have high computational requirements, but the proposed distributed architecture of the services can guarantee this.